\documentclass{article}

\usepackage[utf8]{inputenc}
\usepackage[T1]{fontenc}
\usepackage{amssymb}
\usepackage{amsmath}
\usepackage{amsfonts}
\usepackage{nicefrac}
\usepackage{microtype}
\usepackage{verbatim}
\usepackage{siunitx}
\usepackage{graphicx}
\usepackage{subcaption}
\usepackage{multirow}
\usepackage{booktabs}
\usepackage{soul}
\usepackage{url}

\usepackage{xcolor}
\definecolor{officegreen}{RGB}{0,128,0}
\definecolor{officeblue}{RGB}{0,0,128}
\definecolor{officeburgundy}{RGB}{128,0,64}
\definecolor{resaltarojo}{RGB}{255, 200, 200}
\definecolor{resaltaverde}{HTML}{A0FFC0}
\definecolor{resaltaazul}{RGB}{200, 230, 255}

\usepackage{arxiv} 
\usepackage[colorlinks=true, urlcolor=blue, linkcolor=blue, citecolor=blue]{hyperref}

\title{Temperature in Glass Slides: measurement using Phase Sensitive Optical Coherence Tomography and Computational Modeling}

\author{
    \parbox{\linewidth}{\centering
        Folgueiras J. M.$^{1,2,*}$, Chej G. L.$^{2,3}$, Zurdo L. L.$^{2,3}$, Monastra A. G.$^{2,3}$, Morel E. N.$^{1,2}$, Carusela M. F.$^{2,3}$, Torga J. R.$^{1,2}$
    }
    \vspace{0.3cm} \\
    \normalfont \small $^{1}$Grupo de Fotónica Aplicada, Universidad Tecnológica Nacional -- Facultad Regional Delta (UTN-FRD), Argentina \\
    \normalfont \small $^{2}$Consejo Nacional de Investigaciones Científicas y Técnicas (CONICET), Argentina \\
    \normalfont \small $^{3}$Instituto de Ciencias, Universidad Nacional de General Sarmiento (UNGS), Argentina \\
    \vspace{0.2cm}
    \normalfont \footnotesize $^*$jfolgueiras@frd.utn.edu.ar
}

\date{March 18, 2026}

\begin{document}

\maketitle

%% Abstract
% AFTER Reviewer #3  (to use the past tense uniformly)
\begin{abstract}
Phase-sensitive optical coherence tomography (PhS-OCT) enables precise, contactless measurements of temperature-dependent changes in transparent solids. In this work, we {used} a common-path spectral-domain OCT system to measure optical path differences (OPD) in a 1-mm-thick soda-lime glass slide immersed in a thermal bath. The OPD variation showed a strong linear correlation with temperature in the range of $20-52$~\SI{}{\celsius}, with an experimentally determined sensitivity of $12.4\pm1.9$~[nm/\SI{}{\celsius}]. A theoretical model incorporating the thermo-optic and thermal expansion coefficients of glass was proposed to interpret the measurements, and numerical simulations based on finite volume methods were performed to account for spatial temperature gradients in the system. The simulations {showed} agreement with experimental results within 5\% error, validating the approach. Additionally, repeatability tests using lateral scans at constant temperature demonstrated sub-10 nm stability, supporting future extensions to spatially resolved thermal mapping. This technique provides a low-cost platform for localized temperature sensing in solid transparent materials.
\end{abstract}

\clearpage
\section{Introduction}\label{section:intro_UTN}
Contactless and localized temperature measurement plays a key role in applications that require temperature mapping inside a material, {as well as} the evaluation of spatial and temporal gradients and variations. The use of invasive methods can alter the properties of the material or a complex system, especially in small volumes at the micro and nanoscale, such as in microfluidics, lab-on-a-chip systems \cite{Dos-Reis-Delgado}, or in the manufacturing of semiconductor devices or thin films \cite{Blackburn}.

There are a wide variety of techniques to measure temperatures. Among the most commonly used, we can mention those based on the expansion of liquids and solids, electrical resistance, and radiation \cite{temperaturegeneral}. However, optical methods offer the possibility of measuring contactless, spatially localized, and time-resolved \cite{Mayinger2013opticalmeasurement}. Among the most commonly used optical techniques, we can mention optical fibers, Bragg gratings, infrared thermography, Raman spectroscopy, and interferometric optical methods \cite{Reddy:13}. The latter have gained attention due to their high sensitivity, spatial resolution, and high sampling frequency.

In particular, one of the optical variants, Optical Coherence Tomography (OCT), has been used to obtain temperature measurements through the precise measurement of Optical Path Differences (OPD) at the micron scale, displaying the relationship between temperature changes and the optical and mechanical effects of the sample material \cite{Tsutsumi, volkov2015}.

%%%%%%%%%%%%%%%%%%%%%%%%%%%%%%%%%%%%%%%%%%%%%%%
%%%%%%%%%%%      revisado      %%%%%%%%%%%%%%%%
%%%%%%%  REVIEWER 4  %% comentario 4  %%%%%%%%%
%%%%%%%%%%%%%%%%%%%%%%%%%%%%%%%%%%%%%%%%%%%%%%%

A further development of this technique, known as Phase-Sensitive Optical Coherence Tomography (PhS-OCT)~\cite{zhang2010phase, Zhao2000, Adler:08}, uses phase information from the interference spectrum to overcome the typical axial resolution limitations of standard OCT, improving the sensitivity of the OPD measurement to the nanometer scale. {A common-path configuration of PhS-OCT has shown excellent phase stability, allowing detection of optical path difference changes below 0.5~nm~\cite{lan2017common, li2020ultrahigh}.}

The OPD variation detected by this technique can be correlated with changes in the refractive index and material expansion.{This has enabled a broad range of applications including optical coherence elastography~\cite{li2020ultrahigh}, strain field characterization in polymers~\cite{HUANG2022107566}, monitoring of damage repair in composites~\cite{HUANG2025108689}, curing process characterization~\cite{HUANG2022112184} and temperature sensing.} {In this context,} PhS-OCT has proven to be a highly sensitive technique for measuring temperature and absorption coefficients~\cite{volkov2015, goetz2018interferometric}. It has been successfully used to correlate thermal expansion in applications such as bone surgery~\cite{hamidi2023towards, hamidi2020imaging} and as a complementary modality to photothermal optical coherence tomography (PT-OCT)~\cite{Adler:08}. {More recently, it has been applied to monitor tissue temperature rise and thermal deformation during retinal laser therapy~\cite{Veysset:23,zhuo2025retinal}.}
%%%%%%%%%%%      revisado      %%%%%%%%%%%%%%%%
%%%%%%%%%%%%%%%%%%%%%%%%%%%%%%%%%%%%%%%%%%%%%%%
%%%%%%%  REVIEWER 4  %% comentario 4  %%%%%%%%%
%%%%%%%%%%%%%%%%%%%%%%%%%%%%%%%%%%%%%%%%%%%%%%%

{This work presents a new PhS-OCT approach, specifically optimized for temperature sensing, with core novelty in the use of a standard glass slide as a calibrated interferometric thermometer. To our knowledge, this concept—employing a simple, low-cost, and easily modeled commercial substrate as a temperature probe—has not been previously demonstrated. Our system employs a simplified common‑path configuration in which both surfaces of the slide serve as the sample and reference arms, enabling straightforward implementation. We employ direct empirical calibration of the OPD-temperature relationship through controlled experiments across a physiologically relevant range (20–50°C). Furthermore, we provide comprehensive uncertainty quantification following standardized guidelines, enabling reliable temperature measurements with well-characterized confidence intervals. This end-to-end methodology offers advantages in material simplicity, cost-effectiveness, and minimal invasiveness for applications where introducing complex custom sensors is undesirable.}

We used a Fourier domain optical coherence tomography (FD-OCT) system, a configuration of OCT that employs a spectrometer as the detector, to obtain a relation between the OPD and temperature on a glass slide (soda lime) immersed in a thermal bath.  We {considered} a common-path configuration that provides additional stability against external disturbances \cite{Verma}. 
The results {showed} a linear correlation between OPD and temperature variation \cite{Adler:08}.The capacity of this technique to obtain the temperature inside a liquid was verified using the glass slide as a ``thermometer''.

A computational model of the thermal behavior of the system {was also presented} in this work, incorporating both the thermal expansion of the material and changes in its refractive index. This model {provided} a theoretical framework to support the interpretation of experimental OPD-temperature data. The model was validated with our experimental data, showing good correlation between measurements and simulations. These finite volume-based simulations are widely used for thermal analysis \cite{chej_modeling_2024} and were adapted here to match our experimental conditions.
%%%%%%%%%%%%%%%%%%%%%%%%%%%%%%%%%%%%%%%%%%%%%%%%%%%%%
%%%%%%%%%% 01_INTRODUCCION_UTN %%%%%%%%%%%%%%%%%%%%%%
%%%%%%%%%%%%%     REVISED JOSE        %%%%%%%%%%%%%%%
%%%%%%%%%%%%%%%%%%%%%%%%%%%%%%%%%%%%%%%%%%%%%%%%%%%%%
%%%%%%%%%%%%%%%%%%%%%  END %%%%%%%%%%%%%%%%%%%%%%%%%%
%%%%%%%%%%%%%%%%%%%%%%%%%%%%%%%%%%%%%%%%%%%%%%%%%%%%%

%%%%%%%%%%%%%%%%%%%%%%%%%%%%%%%%%%%%%%%%%%%%%%%%%%%%%
%%%%%%%%%     UTN_THEORETICAL_APPROACH     %%%%%%%%%%
%%%%%%%%%%%%%%%%%%%%%%%%%%%%%%%%%%%%%%%%%%%%%%%%%%%%%

%%%%%%%%%%%%%%%%%%%%%%%%%%%%%%%%%%%%%%%%%%%%%%%%%%%%%
%%%%% 02_UTN_THEORETICAL_APPROACH_cleaned %%%%%%%%%%%
%%%%%%%%%%%%%%%%%%%%%%%%%%%%%%%%%%%%%%%%%%%%%%%%%%%%%
%%%%%%%%%%%%%  REVISED  JOSE %%%%%%%%%%%%%%%% %%%%%%%
%%%%%%%%%%%%%%%%%%%%%%%%%%%%%%%%%%%%%%%%%%%%%%%%%%%%%

\section{Theoretical {Approach } }\label{section:Theoretical_approach_UTN}

To determine the slide's temperature via OPD measurements, we analyzed the interference signal generated between reflections from both surfaces under normal laser incidence. For a laser source with spectral distribution \( S(k) \) (centered at wave vector \( k_0 \)), the intensity $I(k)$ of the interference signal is \cite{Yan:13}:

      \begin{equation}
       \label{eq:intensidad}
%            \begin{align*}
    I(k)=   S(k)\left [(R_{\textrm{R}  }+R_{\textrm{S}}) 
            +2\sqrt{R_{\textrm{R}  }R_{\textrm{S}}} \textrm{ } (cos(2kz(T))\right ]
%            \end{align*}
        \end{equation}
where \( k \) is the wave vector, $R_{\textrm{R}  }$ and $R_{\textrm{s}  }$ are the reflectivities of both surfaces of the slide and $z(T)$ is its OPD at temperature $T$. The phase \( \phi \) of the Fourier transform of the interference signal relates to the OPD through \cite{Yan:13}:

\begin{equation}
\phi(z_i) = k_0 (z_i - 2 z(T))
\label{eq:relacion_lineal}
\end{equation}

Here,  \( z_i \) is the position along the Fourier-conjugate axis of k.
Traditional approaches to obtaining the phase usually employ a single point in the Fourier transform of the interference signal, typically at the maximum of its modulus. In this work, a new method was used, utilizing multiple points within a region named the Maximum Amplitude Zone (MAZ) to improve the accuracy and robustness of the measurement.

The procedure to retrieve the phase {consisted} of the following steps:   

\begin{enumerate}
    \item \textbf{Fourier Transform}: A Fourier transform {was applied} to the interference signal to obtain the modulus and phase in the Fourier-conjugate axis domain (\(z_i\)).
    \item \textbf{Selection of the Maximum Amplitude Zone (MAZ):} A maximum amplitude zone {was defined} around the peak of the modulus, where the signal-to-noise ratio (SNR) is optimal. This zone {was selected} using a threshold of 50\% of the maximum modulus value.
    \item \textbf{Phase Calculation}: The phase {was computed} at multiple points within the MAZ {(33 points centered on the modulus peak)}, instead of using only the central point. This {allowed} averaging the phase values and reducing the impact of noise. {The circular mean of these 33 complex-valued phase samples is taken as the final phase estimate.}
    \item \textbf{OPD Reconstruction}: Based on the computed phase, the OPD {was obtained} using the linear relationship in equation (\ref{eq:relacion_lineal}).
\end{enumerate}

This approach {improved} measurement accuracy by leveraging information from multiple points, in contrast to traditional methods that rely on a single point \cite{Hendargo2011}.
{The key advantage of multi-point averaging is not merely noise reduction, but the effective suppression of phase-wrapping artifacts. In conventional single-point phase retrieval, any measured phase $\phi$ is inherently ambiguous to within $\pm 2\pi$. When computing the phase difference $\Delta\phi$ between two spectra, this ambiguity becomes unbounded—noise-induced jumps at the $\pm \pi$ boundaries can shift $\Delta\phi$ by arbitrary multiples of $2\pi$. By converting each phase sample to a unit complex vector before averaging, this method operates in the complex plane, circumventing discontinuous phase-wrapping artifacts and yielding a stable, unambiguous phase difference.}

This method {integrated} directly with the theoretical approach presented below, where the relationship between OPD and temperature in a material is described. The ability to measure OPD with higher precision allows a more accurate estimation of thermal changes in the material, which is crucial for applications in material characterization and thermal sensing.

% traducido a continuacion
We {proposed} to measure the temperature in a localized region between the two surfaces of a glass slide by integrally leveraging the material’s mechanical and photothermal effects. For this purpose, the experimental setup described in Section \ref{section:experimental} was used. OPD measurements were performed between the two surfaces of the glass slide under controlled temperature variations. A correlation between OPD and temperature was established.

The dependence of the OPD on the sample temperature can be modelled by analysing the mechanical and thermal dynamics. This considers two main contributions: firstly, the volumetric thermal expansion coefficient of the glass slide; and secondly, the variation in the refractive index of the glass slide with temperature (the thermorefractive or thermo-optic effect). Following this model \cite{Adler:08}, the OPD ($z(T)$) between two light beams reflected from each surface of the glass slide at a temperature $T$, is given by:

\begin{equation}
z(T)=L(T)n(T)
    \label{eq:OPD_de_T}
\end{equation}

Here, $L(T)$ is the slide thickness and $n(T)$ is its refractive index. A change in optical path length $\Delta{z}$ that occurs due to a change in temperature $\Delta{T}$ relative to an initial condition, a temperature $T_0$, can be written as:

\begin{equation}
\Delta{z}=z(T_0+\Delta{T})-z(T_0)=L(T_0+\Delta{T}) \; n(T_0+\Delta{T})-L(T_0) \; n(T_0) 
\label{eq:Delta_OPD_de_T}    
\end{equation}

Now it is considered a first-order variation in the refractive index and in the thickness of the slide, with an index variation given by $\frac{dn}{dT}$ and a volumetric coefficient of expansion $\beta$, the change in optical path length can be expressed using:

\begin{align}
L(T_0+\Delta{T}) &= L(T_0)  \times (1+\beta{\Delta{T}}) \nonumber \\  
n(T_0+\Delta{T}) &= n(T_0)+\frac{dn}{dT} \Delta{T}  \nonumber \\
\label{eq:L_cero_y_n_cero}
\end{align} 
The expression for $\Delta{z}$ can be expanded to give:

\begin{equation}
\Delta{z}= \left(L(T_0) \; \frac{dn}{dT} + L(T_0) \beta\;n(T_0) \right) \;{\Delta{T}} + L(T_0) \beta\;\frac{dn}{dT}{\Delta{T}}^2
\label{eq:Delta_OPD_de_T_2_exp_cuadratica}
\end{equation}

%Both mechanical and thermo-optic responses in a soda-lime glass slide were estimated using the model proposed by Ghosh \cite{ghosh1995dispersion}, which provides the thermal expansion coefficient ($\beta$) and thermo-optic coefficient ($dn/dT$) for a simplified ternary system (Na$_2$O--CaO--SiO$_2$). The sample used in this work contains an additional MgO modifier, with a composition of 72.6\% SiO$_2$, 13.0\% Na$_2$O, 8.8\% CaO, and 4.3\% MgO by weight.
 %Ghosh's \textit{Series 2} ((25$-$x)Na$_2$O--xCaO--75SiO$_2$) with $x = 10$ ($\sim$15\% Na$_2$O,  10\% CaO, 75\% SiO$_2$), which lacks MgO, yields $\beta = 8.7 \times 10^{-6}/^\circ$C and $dn/dT = +2.87 \times 10^{-6}/^\circ$C.
{The sample used in this work is composed of 72.6\% SiO$_2$, 13.0\% Na$_2$O, 8.8\% CaO, and 4.3\% MgO by weight.
MgO acts as a network modifier with higher field strength than Na$^+$ or Ca$^{2+}$, reducing both thermal expansion and thermo-optic coefficients. For a similar composition and in the temperature range 20--50$^\circ$C, the thermal expansion coefficient is $\beta = 8.3 \times 10^{-6}/^\circ$C \cite{industrialglasstech}. The thermo-optic coefficient at 800 nm was estimated as $dn/dT = (0.9 \pm 0.3) \times 10^{-6}/^\circ$C. This estimate combines data from NIST Standard Reference Material SRM1822a \cite{NIST_SRM1822a_2008} (which reports $dn/dT = 1.2 \times 10^{-6}$°C$^{-1}$ at 632.8 nm and $2.2 \times 10^{-6}$°C$^{-1}$ at 543.5 nm) with the established trend that $dn/dT$ decreases monotonically with increasing wavelength in soda-lime–silica glasses \cite{ghosh1995dispersion}.}

Assuming these parameters for our sample, the quadratic term in Eq. \eqref{eq:Delta_OPD_de_T_2_exp_cuadratica} becomes negligible, leading to the simplified expression for the optical path difference (OPD) change:

\begin{equation}
\Delta{z} = L(T_0) \; \left(  \; \frac{dn}{dT} + \beta\;n(T_0) \right) \;{\Delta{T}}
\label{eq:Delta_OPD_de_T_lineal}        
\end{equation}

% Modelo empirico de la expansion del vidrio, graficos y procesado en: C:\Users\jfolgueiras\OneDrive - UTN Facultad Regional DELTA\Escritorio\matlab\2022\analisisFASE_extencion_espectro_k\version12\procesado_P4\5_graficar\comparativas

{Applying Eq.~\eqref{eq:Delta_OPD_de_T_lineal} with $L(T_0) = 1$~mm, $n(T_0) = 1.52$, $\beta = 8.3 \times 10^{-6}/^\circ$C, and $dn/dT = 0.9 \times 10^{-6}/^\circ$C yields a theoretical OPD variation rate of $\Delta{z} \approx 13.5$~nm/$^\circ$C. This value shows good agreement with our experimentally measured calibration slope of $12.4 \pm 1.9$~nm/$^\circ$C (Section \ref{section:resultados}).}

%\textcolor{orange}{This analysis highlights a practical consideration when applying optical models to commercial soda-lime glasses, which typically lack manufacturer-provided thermo-optic data.Our empirical calibration approach provides a straightforward alternative while demonstrating reasonable consistency with composition-adjusted theoretical estimates.}

%%%%%%%%%%%%%%%%%%%%%%%%%%%%%%%%%%%%%%%%%%%%%%%%%%%%%
%%%%% 02_UTN_THEORETICAL_APPROACH_cleaned %%%%%%%%%%%
%%%%%%%%%%%%%%%%%%%%%%%%%%%%%%%%%%%%%%%%%%%%%%%%%%%%%
%%%%%%%%%%%%  REVISED  JOSE         %%%%%%%%%%%%%%%%%
%%%%%%%%%%%%%%%%%%%%%%%%%%%%%%%%%%%%%%%%%%%%%%%%%%%%%
%%%%%%%%%%%%%%%%%%%%%  END %%%%%%%%%%%%%%%%%%%%%%%%%%
%%%%%%%%%%%%%%%%%%%%%%%%%%%%%%%%%%%%%%%%%%%%%%%%%%%%%

%%%%%%%%%%%%%%%%%%%%%%%%%%%%%%%%%%%%%%%%%%%%%%%%%%%%%
%%%%%%%%%     UTN_THEORETICAL_APPROACH     %%%%%%%%%%
%%%%%%%%%%%%%%%%%%%%%%%%%%%%%%%%%%%%%%%%%%%%%%%%%%%%%

%%%%%%%%%%%%%%%%%%%%%%%%%%%%%%%%%%%%%%%%%%%%%%%%%%%%%
%%%%%%%%%%     REVISED JOSE     %%%%%%%%%%%%%%%%%%%%%
%%%%%%%%%%%%%%%%%%%%%%%%%%%%%%%%%%%%%%%%%%%%%%%%%%%%%
%%%%%%%%%%     03_EXPERIMENTAL_cleaned     %%%%%%%%%%
%%%%%%%%%%%%%%%%%%%%%%%%%%%%%%%%%%%%%%%%%%%%%%%%%%%%%

\section{Experimental}\label{section:experimental}

\subsection{Setup}\label{subsection:setup}
\begin{figure}[htbp]
    \centering

    \begin{minipage}[t]{0.47\textwidth}
        \centering
        \includegraphics[width=.8\linewidth]{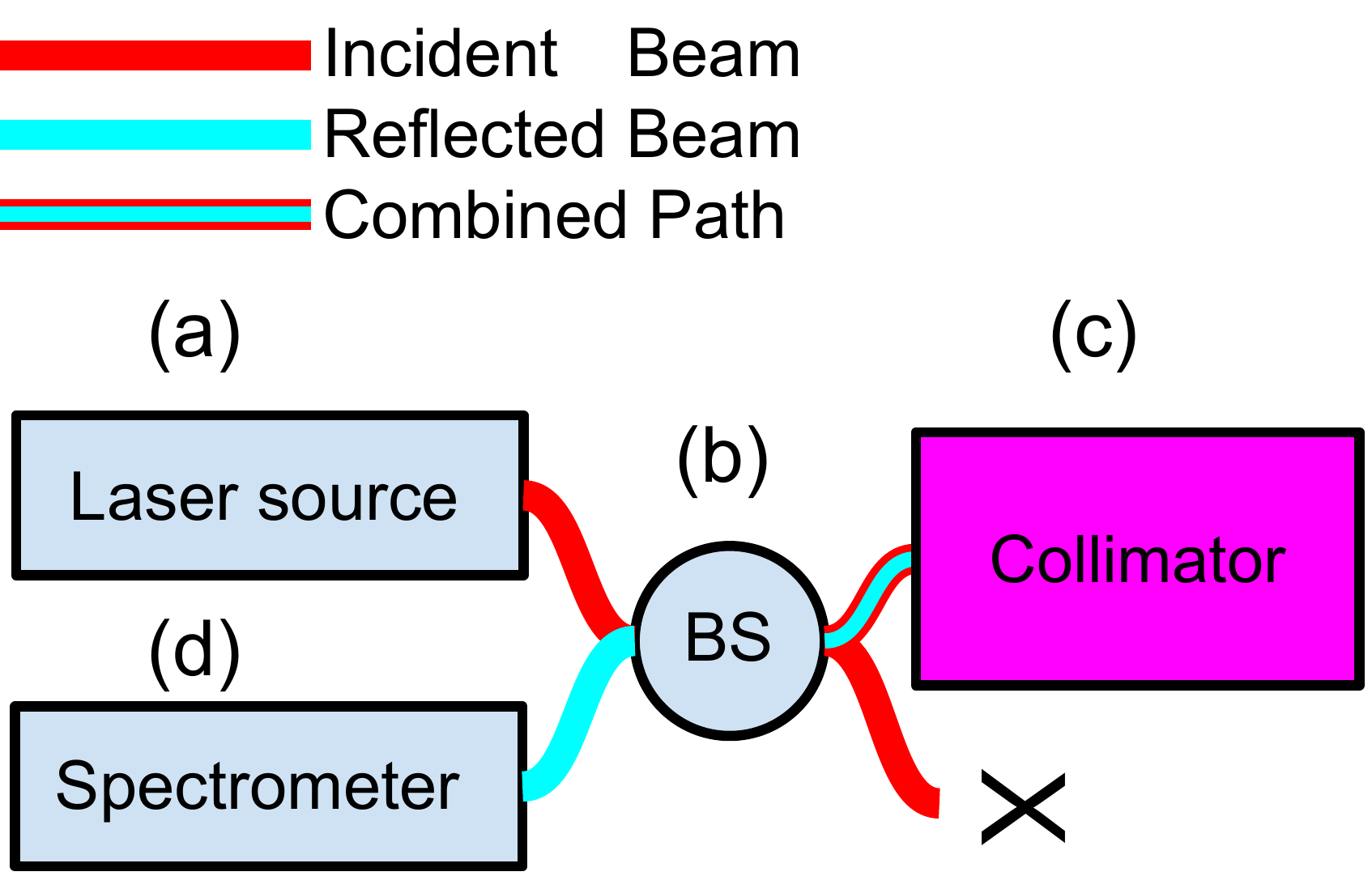}
        \captionof{figure}{Schematic of the low-coherence interferometer setup. Key components: (a) SuperK white-light laser, (b) 50:50 fiber beam splitter, (c) collimator (d) HR4000 spectrometer.}
        \label{fig:esquema_experimental}
    \end{minipage}
    \hfill
    \begin{minipage}[t]{0.47\textwidth}
        \centering
        \includegraphics[width=.8\linewidth]{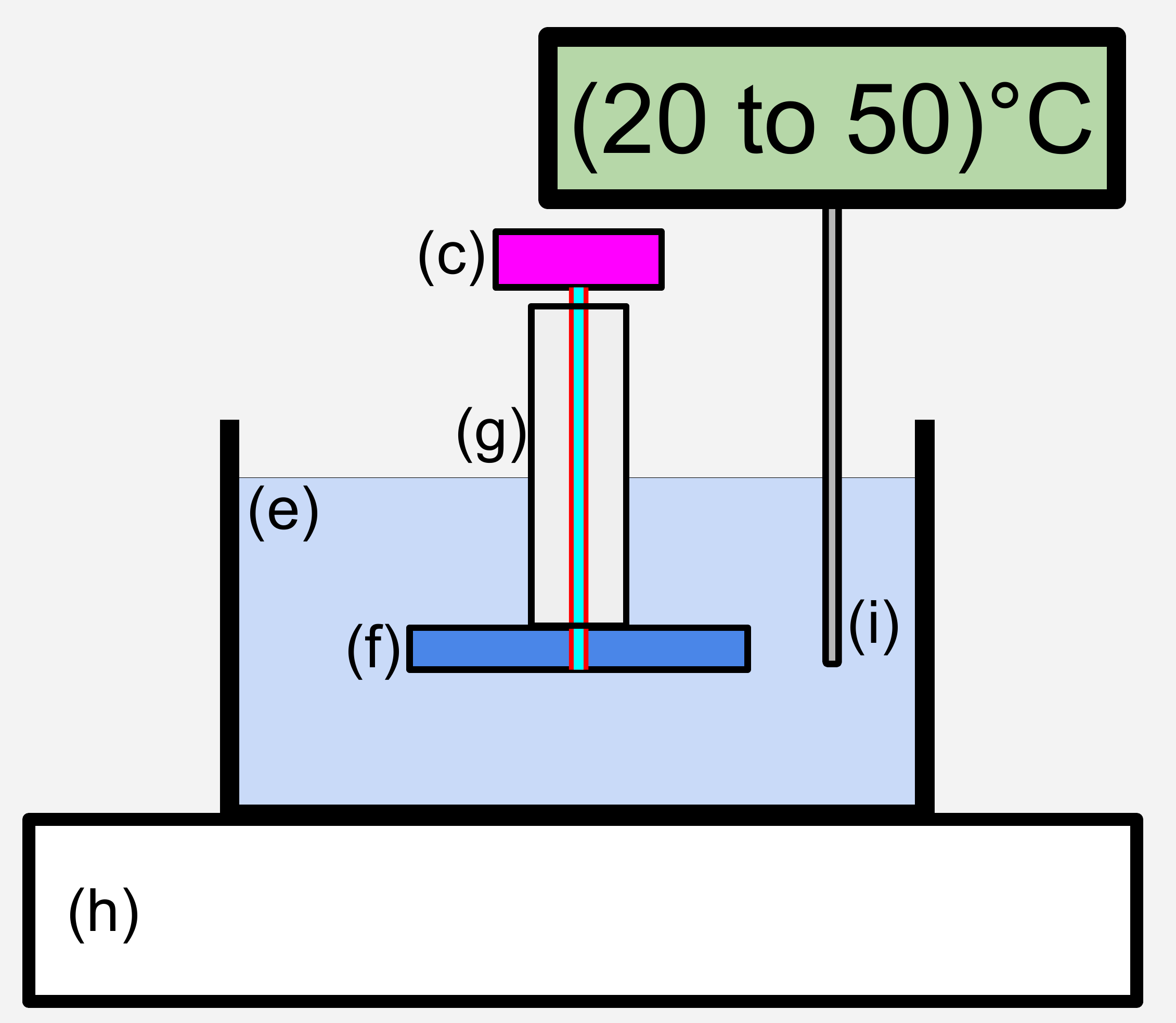}
        \captionof{figure}{Cross-sectional view of the temperature-controlled water bath assembly: (c) collimator (e) glass crystallizer, (f) glass slide, (g) plastic tube maintaining dry optical surface, (h) ceramic hot plate with PID controller, and (i) PT100 thermal sensor.}
        \label{fig:esquema_experimental_tacho_completo}
    \end{minipage}

    \vspace{1em}

    \begin{minipage}[t]{0.47\textwidth}
        \centering
        \includegraphics[width=\linewidth]{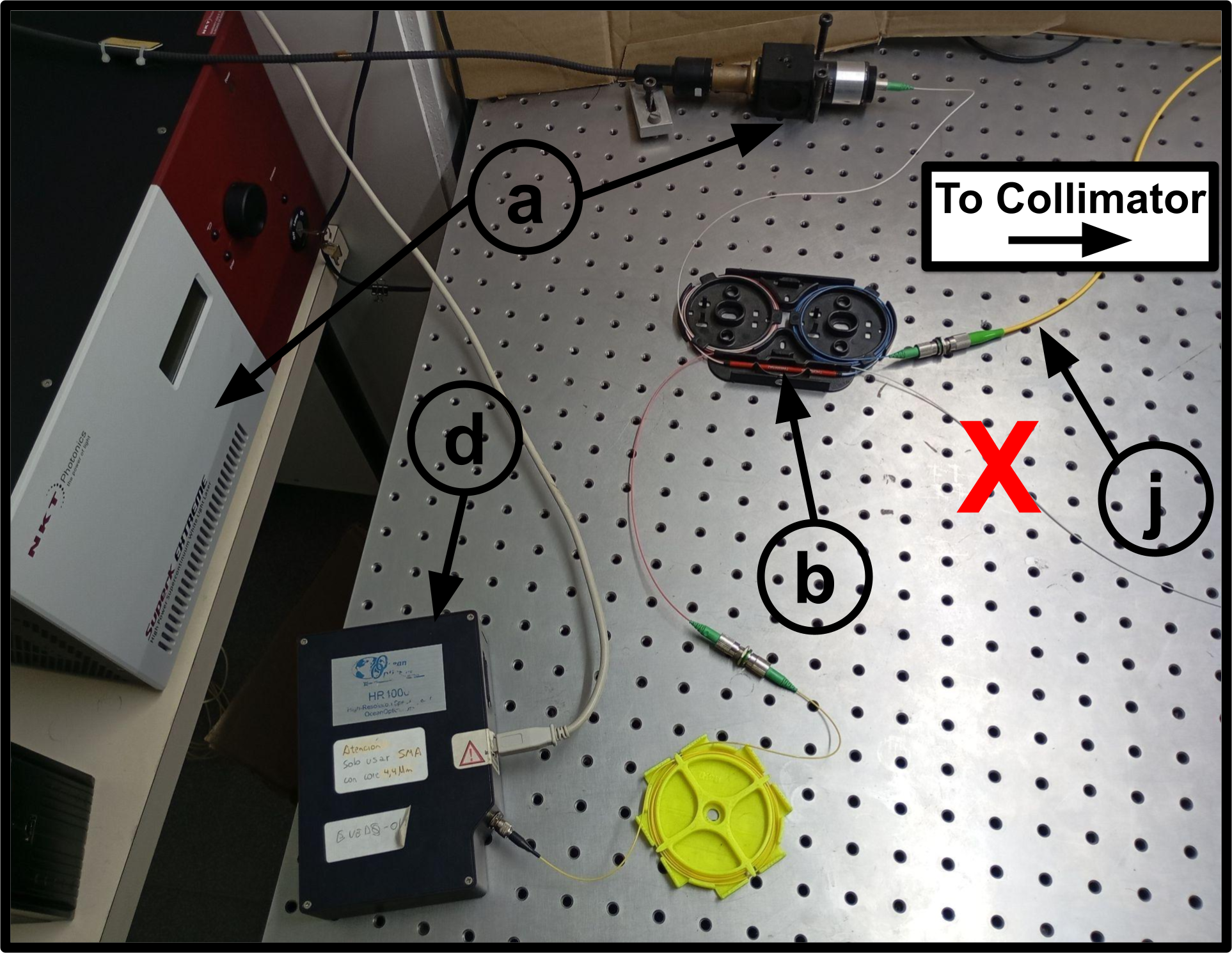}
        \captionof{figure}{Photograph of the experimental setup showing: (a) SuperK white-light laser,(b) 50:50 fiber beam splitter, (d) HR4000 spectrometer, (j) fiber connection from laser source to the collimator.}
        \label{fig:esquema_experimental_foto_completoB}
    \end{minipage}
    \hfill
    \begin{minipage}[t]{0.47\textwidth}
        \centering
        \includegraphics[width=.6\linewidth]{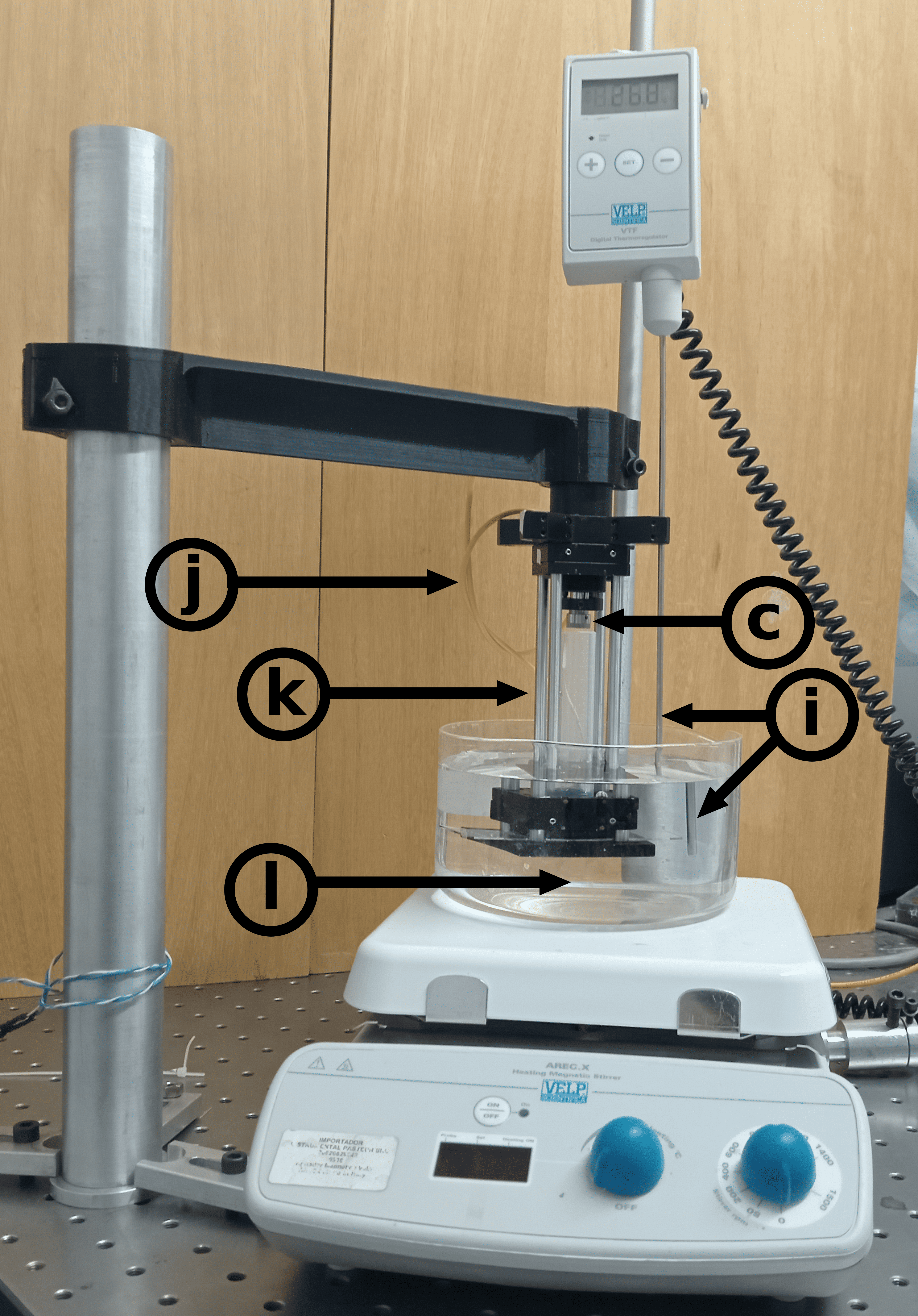}
        \captionof{figure}{Photograph of the experimental setup showing: (c) collimator (i) PT100 thermal sensor. (j) fiber connection from laser source, (k) cage-mounted collimation system, (l) temperature-controlled water bath.}
        \label{fig:esquema_experimental_foto_completo}
    \end{minipage}

\end{figure}

%revised
The experimental setup is shown in Fig.~\ref{fig:esquema_experimental} (optical schematic), Fig.~\ref{fig:esquema_experimental_tacho_completo} (temperature control assembly), and {Fig.~\ref{fig:esquema_experimental_foto_completoB}, Fig.~\ref{fig:esquema_experimental_foto_completo} }(full system photograph).
The light source was a white-light laser (NKT SuperK Extreme EXR4). {The laser was operated at a power level below 5 mW, with its spectral bandwidth restricted to 200 nm by the spectrometer used for detection, comparable to a low-cost superluminescent laser}. The laser output was coupled to a 50:50 fiber splitter, with one arm connected to a fiber-coupled collimator mounted on a tilt-adjustable platform. The collimated beam was aligned for normal incidence on the sample.

 The glass slide used as the sample was suspended inside a crystallizer, filled with distilled water, using an arm attached to the top of the cage system (Fig. \ref{fig:esquema_experimental_foto_completo}). The design allowed complete immersion while keeping a circular area on the top surface (facing the collimator) dry. A small plastic tube was attached to the upper surface of glass slide to prevent water contact, ensuring that the collimated beam interacted only with both slide surfaces.

The crystallizer was placed on a ceramic hot plate with PID temperature control, using a PT100 sensor immersed in the thermal bath near the sample. This system {stabilized} the bath temperature, which simulations confirm to be uniform and representative of the sample conditions.

The light reflected from both surfaces of the glass slide was directed to the spectrometer (Ocean Optics HR4000) for spectral-domain analysis. This configuration eliminated the need for a reference arm since interference occurred exclusively between reflections from the two surfaces of the sample. Once thermal equilibrium was achieved, spectra were recorded over a 5-minute interval at each target temperature. Temperature was monitored simultaneously using the PT100 sensor integrated with the PID controller.

%%%%%%%%%%%%%%%%%%%%%%%%%%%%%%%%%%%%%%%%%%%%%%%%%%%%%
%%%%%%%%%%     03_EXPERIMENTAL_cleaned     %%%%%%%%%%
%%%%%%%%%%%%%%%%%%%%%%%%%%%%%%%%%%%%%%%%%%%%%%%%%%%%%
%%%%%%%%%%     REVISED JOSE     %%%%%%%%%%%%%%%%%%%%%
%%%%%%%%%%%%%%%%%%%%%  END %%%%%%%%%%%%%%%%%%%%%%%%%%
%%%%%%%%%%%%%%%%%%%%%%%%%%%%%%%%%%%%%%%%%%%%%%%%%%%%%

%%%%%%%%%%%%%%%%%%%%%%%%%%%%%%%%%%%%%%%%%%%%%%%%%%%
%%% revision JOSE computational simulations  %%%%%%
%%%%%%%%%%%%%%%%%%%%%%%%%%%%%%%%%%%%%%%%%%%%%%%%%%%
\subsection{Computational simulations}\label{section:Comp_sim}

% Contexto: Breve explicación de la importancia de las simulaciones realizadas dentro del objetivo general del trabajo.
% Objetivo de las simulaciones: Describir qué se busca entender, probar o validar con las simulaciones.

{Temperature calibration requires system heating, but localized heat sources create non-uniform thermal baths. These produce gradients and fluctuations that affect local temperature measurements. }

{According to Rayleigh–Bénard convection theory, the thermophysical properties of water are expected to promote rapid thermal equilibration of the systems, thereby minimizing temperature gradients between the glass substrate and the thermal sensor. For Rayleigh numbers ($Ra$) exceeding 1000—as is the case in our study—the formation of convection cells is assured, resulting in a homogenized temperature distribution throughout the water bath. The Rayleigh number is defined by the relation:}

\begin{equation}
    Ra = \frac{g \beta}{\nu \alpha}(T_b - T_{\infty})L^3    
\end{equation}

{where $g$ denotes gravitational acceleration, $\beta$ is the thermal expansion coefficient, $\nu$ the kinematic viscosity, and $\alpha$ the thermal diffusivity of the fluid. Here, $T_b$ represents the base temperature, $T_{\infty}$ the ambient temperature at the top, and $L$ the vertical height of the fluid column. To more rigorously investigate the thermal behavior of the system—particularly during the transient regime—we conducted a series of high-fidelity numerical simulations. By mapping the thermal distribution, the numerical simulations {aimed to validate the absence of discrepancy} between: (1) PT100 sensor readings and (2) actual glass slide temperature at the OPD measurement point.}

{We performed numerical simulations using OpenFOAM \cite{openfoam} to solve the conjugate heat transfer problem via the Finite Volume Method (FVM). These simulations served two key purposes: (1) characterizing the fluid (water) behavior and temperature distribution in our experimental setup, and (2) providing independent temperature validation.}

\begin{figure}[ht]
    \centering
    \includegraphics[width=0.7\linewidth]{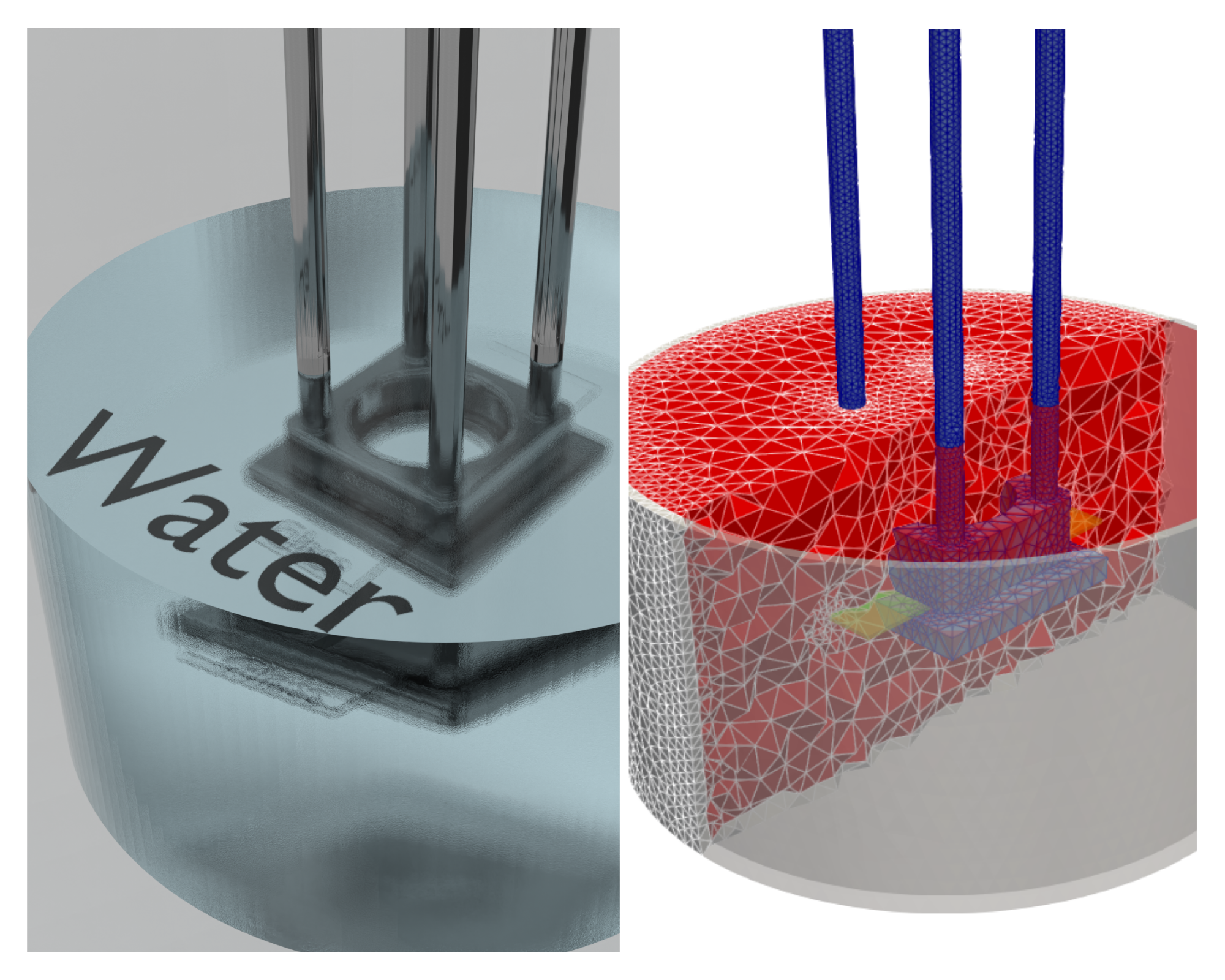}
    \caption{{3D computational model {(left)} and mesh {(right)} of the experimental setup simulated in OpenFOAM}. {The colors on the right serve as a guide to distinguish the different regions of the computational model: fluid (red), slide glass (yellow) and cage-mounted collimation system (blue).}}
    \label{fig:dispositivo_simulado}
\end{figure}

% ----------------------------------------------------------------------------------------
% Descripción del dominio y malla --------------------------------------------------------
% Geometría del dominio: Explicar la geometría del modelo utilizado.
{The computational domain (Fig. \ref{fig:dispositivo_simulado}) precisely {replicated} the experimental geometry, including: an aluminum stabilization stage in direct contact with the glass slide, a water bath ensuring thermal equilibrium during measurements and the complete thermal boundary conditions of the system.}

We {considered} a three-dimensional mesh model (Fig. \ref{fig:dispositivo_simulado}) that was generated using \texttt{Salome Meca}. Primarily employing tetrahedral volumes, our goal was to achieve the best possible results by refining the mesh at the surface {region} interfaces with a minimum volume of $4.18027\times10^{-10}m^3$, a key factor to capture with good accuracy the underlying phenomena.

For the water confined in the crystalliser, a laminar regime {was considered}, with the thermophysical properties of the fluid varying as a function of temperature.

For the boundary conditions, {we considered} free air convection plus radiation at the crystallizer external walls as described by Eq. \ref{eq:ecuacion_contorno_vaso_vidrio}. The same boundary condition applies to the top water surface in contact with the air room. However, the boundary condition on the crystallizer floor {was} the same temperature {measured} in the real experiment, which {increased} by 5°C every 5,000 seconds. This temperature {was} time-dependent and {started} at 292.35K, {reached} 337K, and then {decreased} to 312K.

\begin{equation}
    q_{tot} = h_{tot}(T_{\infty} - T_{wall}(t))
    \label{eq:ecuacion_contorno_vaso_vidrio}
\end{equation}

Where $T_\infty = 292.35K$ and $h_{tot}=15W/m^2K$. This $h_{tot}$ {is composed of} $h_{conv}=10W/m^2K$ convection plus $h_{rad}=5W/m^2K$ radiation \cite{chej_modeling_2024}.

The whole physical domain {had} an initial temperature of $T_{\infty}$. The initial velocity {of the} fluid was zero, and the initial pressure was atmospheric pressure.

%%%%%%%%%%%%%%%%%%%%%%%%%%%%%%%%%%%%%%%%%%%%%%%%%%%
%%% revision JOSE computational simulations  %%%%%%
%%%%%%%%%%%%%%%%%%%%%%%%%%%%%%%%%%%%%%%%%%%%%%%%%%%%

%%%%%%%%%%%%%%%%%%%%%%%%%%%%%%%%%%%%%%%%%%
%%%%%%%    revisada JOSE %%%%%%
%%%%%%%%%%%%%%%%%%%%%%%%%%%%%%%%%%%%%%%%%%%
%%%%%%%%%%     04_MEASUREMENTS_cleaned     %%%%%%%%%%
%%%%%%%%%%%%%%%%%%%%%%%%%%%%%%%%%%%%%%%%%%%%%%%%%%%%%
\subsection{Measurements and Data Processing} \label{Mediciones_Procesado}

The experiments were carried out using a soda lime glass slide \cite{Marienfeld_Slide} (third hydrolytic class) as the sample. The {choice} was made to show that this technique can be implemented in a low-cost, widely available element that can be easily adapted to specific places and shapes.
OPD measurements were performed using the interference signal between the slide faces, at a controlled temperature ranging from \SI{20}{\celsius} to \SI{50}{\celsius}.

\begin{figure}[hbtp]
    \centering
    \includegraphics[width=0.65\linewidth]{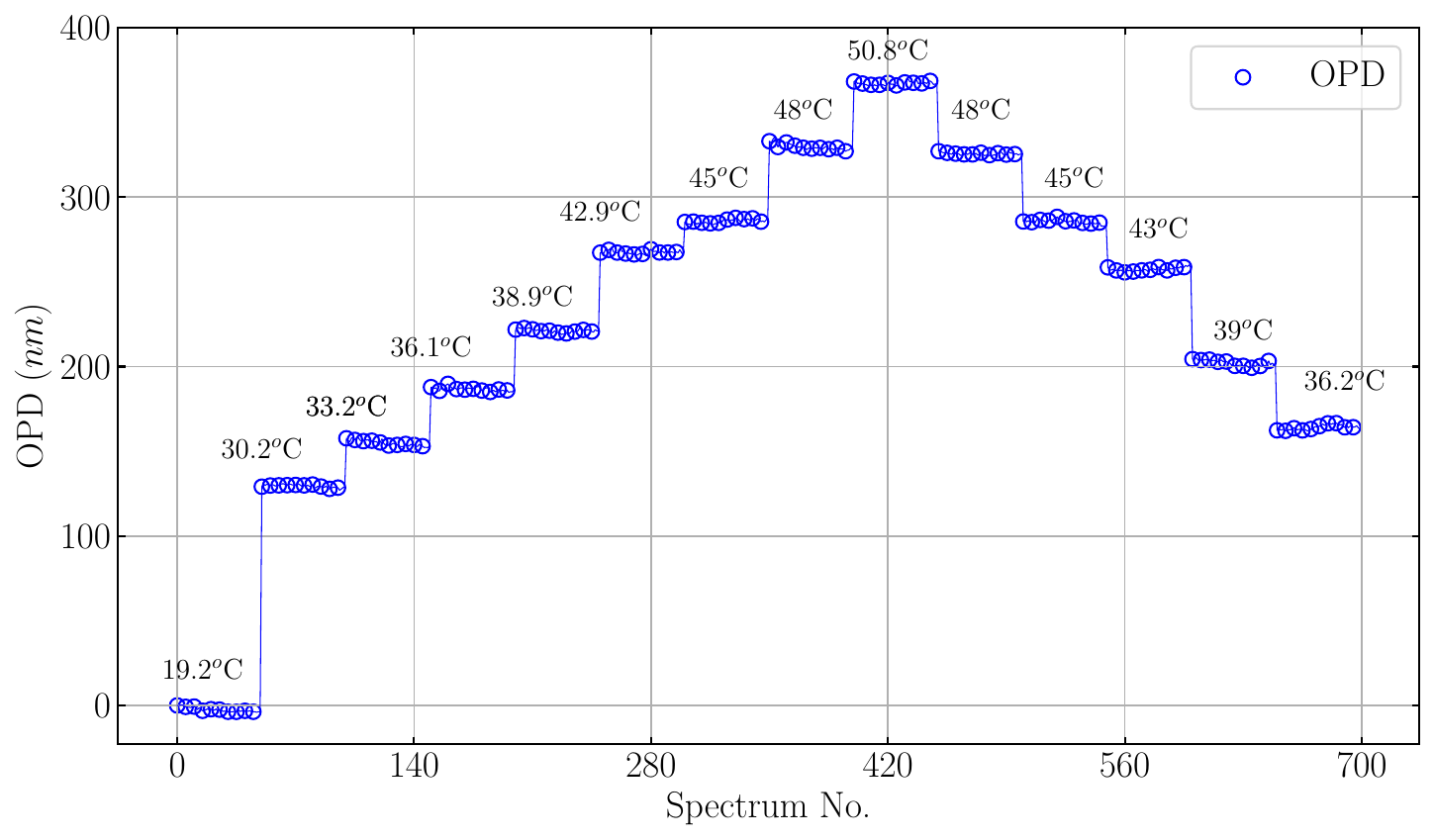}
    \caption{OPD measurements during steady-state conditions (Series S5). Each plateau corresponds to a stable temperature phase.{Temperature labels indicate PT100 sensor readings.}} 
    \label{fig:escalones}
\end{figure}

\begin{figure}[hbtp]
    \centering
    \includegraphics[width=0.7\linewidth]{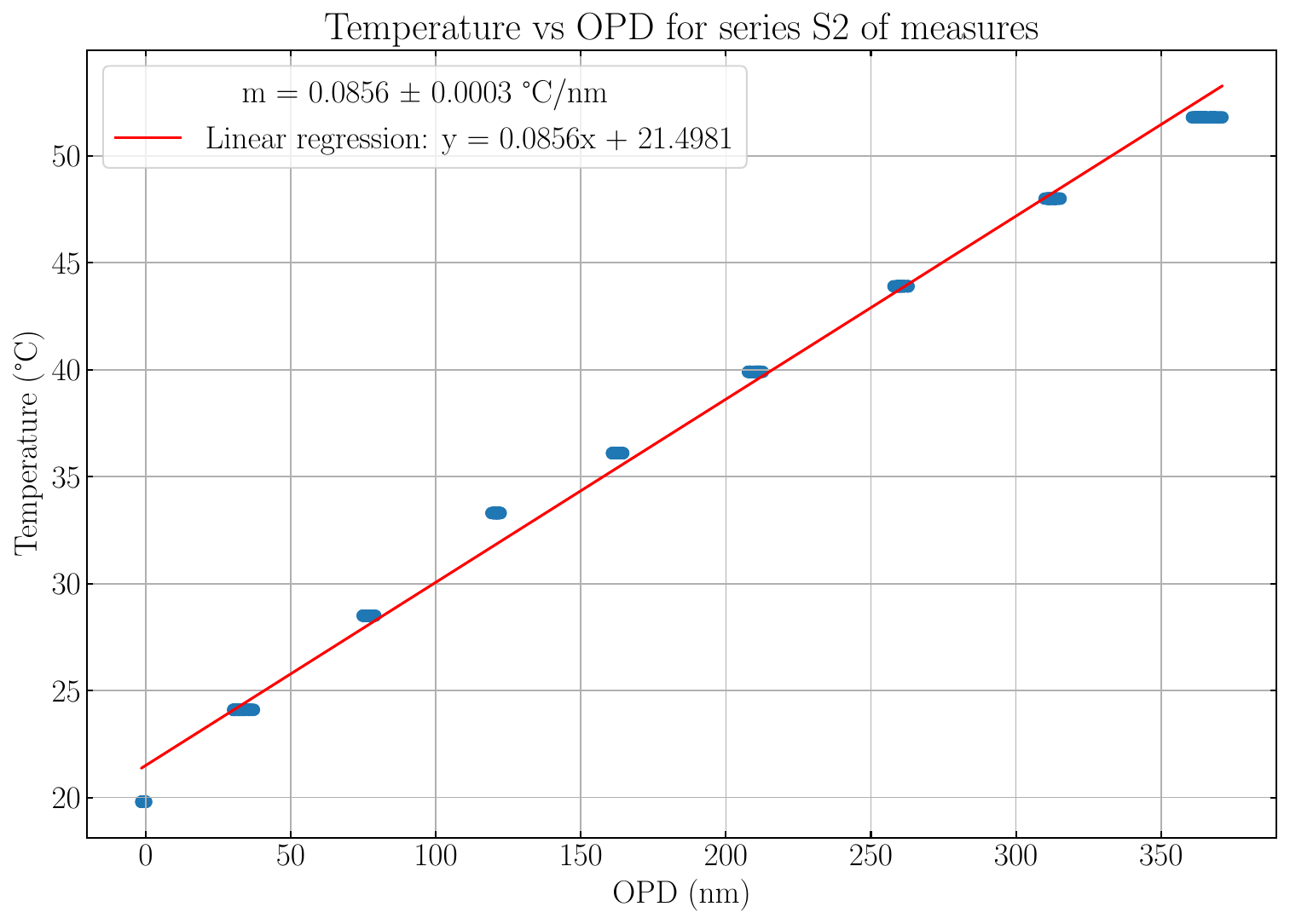}
    \caption{Linear regression of OPD vs. temperature for S2 Series  (slope = ($85.6 \pm 0.3) \cdot10^{-3} $ °C/nm, $R^2 = 0.9929$.}
    \label{fig:linear trend}
\end{figure}

For each target temperature, the sample was thermally stabilized before data acquisition, using the experimental setup described in Section~\ref{subsection:setup}. Spectra were acquired over a 5-minute interval at each temperature value, and the OPD was determined using the PhS-OCT technique described in Section \ref{section:Theoretical_approach_UTN}. The OPD variation exhibited a step-like profile (Fig.~\ref{fig:escalones}), where each step corresponded to a stable thermal phase at a constant temperature. Multiple step series were measured and analyzed through linear regression, revealing a consistent linear response (Fig.~\ref{fig:linear trend}). The experimental data agreed with the theoretical OPD-temperature dependence predicted by Eq.~\eqref{eq:Delta_OPD_de_T_lineal} and are presented in Section~\ref{section:resultados}.

%%%%%%%%%%%%%%%%%%%%%%%%%%%%%%%%%%%%%%%%
%%%%%%  revisada JOSE
%%%%%%%%%%%%%%%%%%%%%%%%%%%%%%%%%%%%%%%

%\caption{\hl{Linear fit applied to the S2 series measurement steps, with OPD (in nm) on the x-axis and Temperature on the y-axis.}}

%%%%%%%%%%%%%%%%%%%%%%%%%%%%%%%%%%%%%%%%%%%%%%%%%%%%%
%%%%%%%%%%     04_MEASUREMENTS_cleaned     %%%%%%%%%%
%%%%%%%%%%%%%%%%%%%%%%%%%%%%%%%%%%%%%%%%%%%%%%%%%%%%%
%%%%%%%%%%%%%%%%%%%%%  END %%%%%%%%%%%%%%%%%%%%%%%%%%
%%%%%%%%%%%%%%%%%%%%%%%%%%%%%%%%%%%%%%%%%%%%%%%%%%%%%

%%%%%%%%%%%%%%%%%%%%%%%%%%%%%%%%%%%%%%%%%%%%%%%%%%%%
%%%%%%%%%%%        RESULTADOS       %%%%%%%%%%%%%%%% 
%%%%%%%%%%%%%%%%%%%%%%%%%%%%%%%%%%%%%%%%%%%%%%%%%%%%

%%%%%%%%%%%%%%%%%%%%%%%%%%%%%%%%%%%%%%%%%%%%%%%%%%%%%
%%%%%%%%%%%%       REVISADO JOSE      %%%%%%%%%%%%%%%
%%%%%%%%%%%%%%%%%%%%%%%%%%%%%%%%%%%%%%%%%%%%%%%%%%%%%
%%%%%%%%%   05_RESULTADOS_MEDIDAS_cleaned   %%%%%%%%%
%%%%%%%%%%%%%%%%%%%%%%%%%%%%%%%%%%%%%%%%%%%%%%%%%%%%%

\section{Results}\label{section:resultados}

%\subsection{Statistical study of experimental results}
We conducted multiple independent measurement series (labeled S1, S2, S3, etc.), as documented in Table~\ref{tabla_mediciones_series1}. Each series followed an identical protocol: first stabilizing the system at discrete target temperatures (e.g., 25°C, 30°C), then acquiring spectra over a 5-minute period at each equilibrium temperature, and finally performing linear regression analysis on the complete dataset. All series demonstrated robust linear OPD-temperature dependence, with correlation coefficients ($R^2$) consistently exceeding 0.99.
%%%%%%%%%%%%%%%%%%%%%%%%%%%%%%%%%%%%%%%%%%%%
%%%%%%%%%%%%%%      TABLA         %%%%%%%%%%
%%%%%%%%%%%%%%%%%%%%%%%%%%%%%%%%%%%%%%%%%%%%

%%%%%%%%%%%%%%%%%%%%%%%%%%%%%%%%%%%%%%%%%%%%%%%%%%%
%%%%%%%%%%% Tabla V2 (compacta) %%%%%%%%%%%%%%%%%%%
%%%%%%%%%%%%%%%%%%%%%%%%%%%%%%%%%%%%%%%%%%%%%%%%%%%

\begin{table}[ht]
\centering
\begin{tabular}{|c|c|c|c|c|}
\hline
\rule{0pt}{25pt} % Espacio vertical adicional
\shortstack{Steps\\Series} & Steps & \shortstack{Temperatures\\(°C)} & \shortstack{Slope\\(°C/nm)} & \shortstack{Sensitivity\\(nm/°C)} \\[2pt]
\hline
S1 &  5 & 22.5-37.5 & $86.6 \times 10^{-3}$ & 11.6 \\
S2 &  9 & 20-52.5 & $85.6 \times 10^{-3}$ & 11.7 \\ 
S3 &  9 & 20-51 & $86.7 \times 10^{-3}$ & 11.5 \\
S4 &  6 & 36-51 & $72.2 \times 10^{-3}$ & 13.9 \\
S5 &  9 & 21-52 & $81.5 \times 10^{-3}$ & 12.3 \\
S6 &  4 & 39-52 & $76.5 \times 10^{-3}$ & 13.1 \\
S7 &  5 & 20-33 & $74.5 \times 10^{-3}$ & 13.4 \\
\hline 
\end{tabular}
\caption{Measurement series summary showing experimental parameters and thermal sensitivity. The added column converts slope values to more intuitive sensitivity units~(nm/°C).}
\label{tabla_mediciones_series1}
\end{table}

%%%%%%%%%%%%%%%%%%%%%%%%%%%%%%%%%%%%%%%%%%%%%%%%%%%
%%%%%%%%%%%%%%%%%%%%%%%%%%%%%%%

The OPD-temperature sensitivity was determined as $12.4 \pm 1.9$ [nm/\SI{}{\celsius}] (mean $\pm 2\sigma$), with the error propagated from the linear fit slope from Table~\ref{tabla_mediciones_series1} ($(81 \pm 12)\times 10^{-3}$ [\SI{}{\celsius}/nm])

%\subsection{Simulation results vs experimental results}

{Figure~\ref{fig:simulation-results} shows the exponential approximation for each temperature plateau (maintained for 5,000 seconds) {and presents the transient thermal response of the system}. The simulation, performed using Series S3 (upward) and S4 (downward) data, demonstrates good agreement with experiments. }
 
{As mentioned in Section~\ref{section:Comp_sim}, the 3D computational model was used to assess the thermal homogeneity of the water bath and to validate the correlation between the PT100 sensor readings and the actual temperature at the glass slide's Optical Path Difference (OPD) measurement point. The simulation confirmed temperature uniformity due to the water bath in which the glass plate was immersed. At the peak heating point (Figure \ref{fig:simulation-results} at 10.5 hs), the results indicated a temperature difference of $0.15^{\circ}$C between the center of the glass plate and the water cell corresponding to the experimental sensor location.}

{{The simulation captured the thermal inertia and the stepwise heating/cooling stages. To quantify the agreement between the experimental data ($T_{\text{exp}}$) and numerical predictions ($T_{\text{sim}}$), the Mean Absolute Percentage Error (MAPE) was calculated as follows:
\begin{equation}
    \text{MAPE} = \frac{100\%}{n} \sum_{i=1}^{n} \left| \frac{T_{\text{exp},i} - T_{\text{sim},i}}{T_{\text{exp},i}} \right|
\end{equation}
Where $n$ is the number of total plateaus corresponding to a stable temperature phase of Figure \ref{fig:simulation-results}. $T_{exp}$ and $T_{sim}$ are the water temperature when the plateau reaches a steady state. Considering the temperature values in the Celsius scale, the MAPE for the entire series S3 (up-ward) and S4 (downward) data was $4.97\%$. Along the series, the maximum temperature deviation was $5.65^o$C.
This discrepancy, while moderate in percentage terms, is significant relative to the working range and warrants a more detailed interpretation. Three main factors contribute to the observed differences, particularly at elevated temperatures.}

{First, the experimental setup employs a PID-controlled heater. The simulation simplifies the heater behavior by imposing a recorded base temperature as a Dirichlet boundary condition, assuming perfect and continuous temperature maintenance.
Second, the model assumes perfect thermal contact between the heater plate and the crystallizer bottom, with flat, fully contacting surfaces.
Third, the simulation employs a constant combined heat transfer coefficient $h_{\text{tot}}$ (Eq.~\eqref{eq:ecuacion_contorno_vaso_vidrio}) corresponding to an idealized Rayleigh-Bénard configuration with perfectly quiescent ambient conditions.}
{However, the numerical analysis confirms the validity of our experimental assumptions. The simulation shows that despite the localized heating at the base, the Rayleigh-Bénard convection cells effectively homogenize the water bath.}

\begin{figure}[ht]
    \centering
    \includegraphics[width=0.7\linewidth]{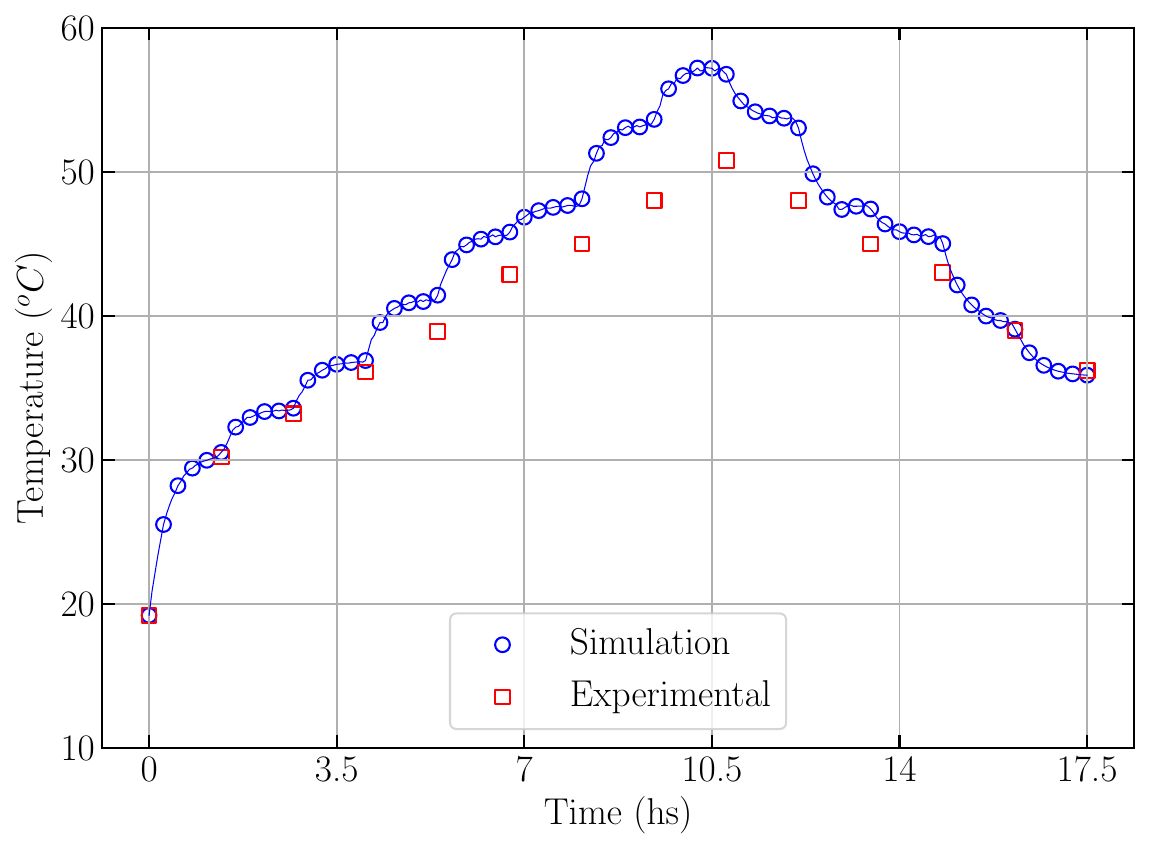}
    \caption{{Comparison of experimental {(PT$100$ sensor measurements - red open squares)} and simulated {(open blue circles)} water temperature. Simulation based on Series S3 (upward) and S4 (downward) data.}}
    \label{fig:simulation-results}
\end{figure}

%%%%%%%%%%%%%%%%%%%%%%% UTN UTN UTN UTN UTN UTN %%%%%%%%%%%%%%%%%%%%%%%
%\subsection{Repeatability and Stability of Profile Scans}  

%%%%%%%%%%%%%%%%%%%%%%%%%%%%%%%%%%%%%%%%%%%%%%%
%%%%%%%%    uncertainty analysis  %%%%%%%%%%%%%%
%%%%%%%%%%%%%%%%%%%%%%%%%%%%%%%%%%%%%%%%%%%%%%%
\subsection{Measurement Uncertainty Evaluation}
\label{subsec:incertidumbre}
An uncertainty analysis was performed to estimate the uncertainty associated with the temperature measurements obtained with the proposed system. The main contributors considered were: (1) OPD repeatability (including spectrometer phase noise and mechanical stability), (2) the uncertainty in the calibration slope \(m\), and (3) the uncertainty of the reference platinum thermometer (PT100), the latter being excluded from the uncertainty evaluation of the OPD-based temperature measurement.
\subsubsection*{Experimental Repeatability}
OPD repeatability was evaluated by processing five independent series of 1000 measurements each, using the setup described in Section~\ref{subsection:setup}, with the ceramic hot plate turned off and under stable ambient temperature conditions (PT100, \(\pm 0.5\,^\circ\mathrm{C}\)). The experimental standard deviation was \(s=\SI{1.4}{\nano\meter}\), which was taken as the standard uncertainty of the OPD.
\subsubsection*{Measurement Model and Propagation Errors}
The temperature difference is obtained as \(T-T_{\text{ref}}=\mathrm{OPD}/m\), with \(m=\SI{12.4}{\nano\meter\per\degreeCelsius}\). The combined standard uncertainty \(u_c(T-T_{\text{ref}})\) was calculated by propagation in quadrature assuming uncorrelated variables.

The relative combined standard uncertainty decreases from $\sim$32\% for \SI{5}{\nano\meter} (\SI{0.4}{\degreeCelsius}) to $\sim$16\% for \SI{50}{\nano\meter} (\SI{2.5}{\degreeCelsius}), as the uncertainty associated with \(m\) becomes dominant over the fixed OPD contribution. For a temperature variation of \SI{4}{\degreeCelsius}, the combined standard uncertainty of \(T-T_{\text{ref}}\) is \SI{\pm0.6}{\degreeCelsius}.    

%%%%%%%%%%%%%%%%%%%%%%%%%%%%%%%%%%%%%%%%%%%%%%%
%%%%%%%%    uncertainty analysis  %%%%%%%%%%%%%%
%%%%%%%%%%%%%%%%%%%%%%%%%%%%%%%%%%%%%%%%%%%%%%%

\subsection{Spatial Profiling Capability and Repeatability}
\label{subsec:spatial_profiling}
{To characterize the system's spatial profiling capability, we performed repeated linear scans along identical trajectories on the glass slide under constant temperature conditions (room temperature). The measured OPD profiles demonstrate consistent sub-10 nm repeatability across 11 independent scans (M0-M10 in Fig.~\ref{fig:repeatability_a}), with residual noise between scans (M0-M1 difference in Fig.~\ref{fig:repeatability_b}) confirming this precision. The observed stability enables detection of potential thermal gradients ($\Delta T/\Delta x$) beyond single-point measurement limitations.}

\subsection{Spatial Averaging Benefits}
\label{subsec:averaging_benefits}
{While the axial resolution reaches the nanometer scale, the laser spot (on the order of micrometers) provides intrinsic spatial averaging that suppresses small-scale thermal fluctuations. This occurs through statistical cancellation of forward/backward displacements within the illuminated area, enhancing stability for bulk temperature measurements. The effect is particularly advantageous when assessing intensive properties where local variations should not dominate global measurements.}

\begin{figure}[ht]
    \centering
    \begin{subfigure}[b]{0.49\linewidth}
        \centering
        \includegraphics[width=\linewidth]{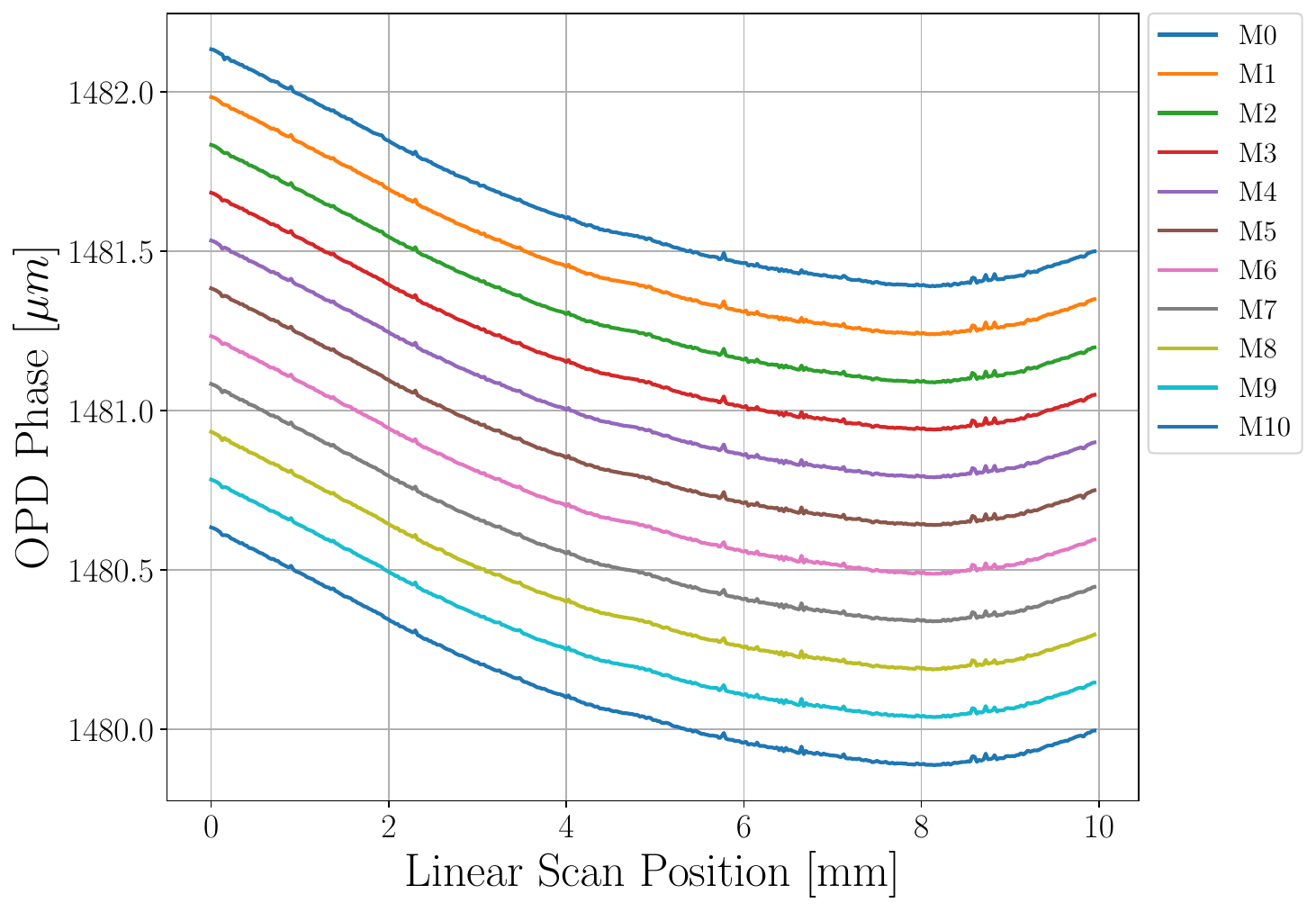}
        \caption{OPD profiles for 11 repeated scans (M0-M10) along identical trajectories. Vertical offsets separate curves for visualization.}        
        \label{fig:repeatability_a}
    \end{subfigure}
    \hfill
    \begin{subfigure}[b]{0.41\linewidth}
        \centering
        \includegraphics[width=\linewidth]{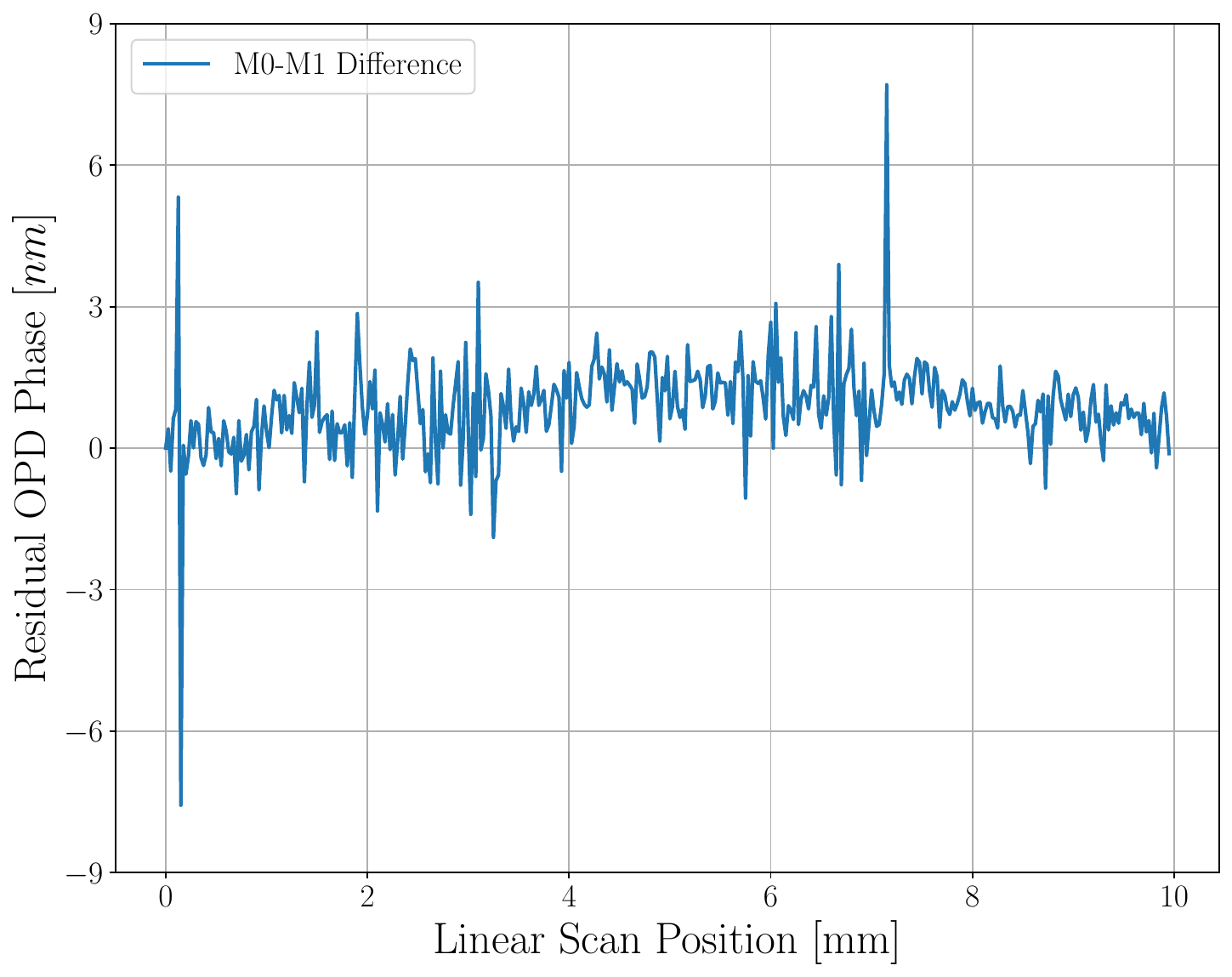}
        \caption{Residual noise (M0-M1 difference) demonstrating $\leq$10 nm repeatability.}
        \label{fig:repeatability_b}
    \end{subfigure}
    \caption{Repeatability analysis of OPD measurements showing (a) superimposed scan profiles {(OPD includes refractive-index factor, \(n \approx 1.52\))} and (b) inter-scan consistency.}
    \label{fig:repeatability}
\end{figure}

%%%%%%%%%%%%%%%%%%%%%%%%%%%%%%%%%%%%%%%%%%%%%%%%%%%%%
%%%%%%%%%   05_RESULTADOS_MEDIDAS_cleaned   %%%%%%%%%
%%%%%%%%%%%%%%%%%%%%%%%%%%%%%%%%%%%%%%%%%%%%%%%%%%%%%
%%%%%%%%%%%%%%%%%%%%%  END %%%%%%%%%%%%%%%%%%%%%%%%%%
%%%%%%%%%%%%%%%%%%%%%%%%%%%%%%%%%%%%%%%%%%%%%%%%%%%%%
%%%%%%%%%%%%       REVISADO JOSE       %%%%%%%%%%%%%%
%%%%%%%%%%%%%%%%      END      %%%%%%%%%%%%%%%%%%%%%%
%%%%%%%%%%%%%%%%%%%%%%%%%%%%%%%%%%%%%%%%%%%%%%%%%%%%%

%%%%%%%%%%%%%%%%%%%%%%%%%%%%%%%%%%%%%%%%%%%%%%%%%%%%
%%%%%%%%%%%        RESULTADOS       %%%%%%%%%%%%%%%% 
%%%%%%%%%%%%%%%%%%%%%%%%%%%%%%%%%%%%%%%%%%%%%%%%%%%%

%%%%%%%%%%%%%%%%%%%%%%%%%%%%%%%%%%%%%%%%%%%%%%%%%%%%%
%%%%%%%%%%           CONCLUSIONES          %%%%%%%%%%
%%%%%%%%%%%%%%%%%%%%%%%%%%%%%%%%%%%%%%%%%%%%%%%%%%%%%

%%%%%%%%%%%%%%%%%%%%%%%%%%%%%%%%%%%%%%%%%%%%%%%%%%%%%
%%%%%%%%%%%%       REVISADO JOSE      %%%%%%%%%%%%%%%
%%%%%%%%%%%%%%%%%%%%%%%%%%%%%%%%%%%%%%%%%%%%%%%%%%%%%
%%%%%%%%%%    06D_CONCLUSIONES_cleaned     %%%%%%%%%%
%%%%%%%%%%%%%%%%%%%%%%%%%%%%%%%%%%%%%%%%%%%%%%%%%%%%%
\section{Conclusions}

{In this work, we {proposed} a phase-sensitive optical coherence tomography method for contactless temperature measurement through optical path difference monitoring. Our approach achieved three key advances: First, we demonstrated $12.4 \pm 1.9$~nm/\SI{}{\celsius} sensitivity in 1-mm-thick soda-lime glass, with computational simulations validating the measurements within {5\% MAPE} error margins. 
{Second, we refined the theoretical OPD-temperature model to account for the specific composition of our glass substrate. Incorporating literature values for MgO-containing soda-lime glasses ($\beta = 8.3\times 10^{-6}/^\circ$C) and estimating the thermo-optic coefficient at 800 nm ($dn/dT = 0.9\times 10^{-6}/^\circ$C) yielded a theoretical value of $\sim$13.5~nm/\SI{}{\celsius}. This estimate shows 8.5\%  agreement with our experimental result. The analysis highlights the practical challenge of applying established optical models to commercial substrates lacking manufacturer-provided thermo-optic data.}
{Third, we established sub-10 nm repeatability in spatial profiling scans, a critical prerequisite for thermal gradient mapping applications.}

{These findings create new opportunities in three domains: low-cost thermal sensing for industrial process monitoring, biomedical temperature mapping, and composition-specific calibration of thermo-optic materials. Looking ahead, we will focus on implementing 2D thermal gradient reconstruction and expanding validation to diverse glass compositions.}

%%%%%%%%%%%%%%%%%%%%%%%%%%%%%%%%%%%%%%%%%%%%%%%%%%%%%
%%%%%%%%%%    06D_CONCLUSIONES_cleaned     %%%%%%%%%%
%%%%%%%%%%%%%%%%%%%%%%%%%%%%%%%%%%%%%%%%%%%%%%%%%%%%%
%%%%%%%%%%%%       REVISADO JOSE       %%%%%%%%%%%%%%
%%%%%%%%%%%%%%%%      END      %%%%%%%%%%%%%%%%%%%%%%
%%%%%%%%%%%%%%%%%%%%%%%%%%%%%%%%%%%%%%%%%%%%%%%%%%%%%

%%%%%%%%%%%%%%%%%%%%%%%%%%%%%%%%%%%%%%%%%%%%%%%%%%%%%
%%%%%%%%%%           CONCLUSIONES          %%%%%%%%%%
%%%%%%%%%%%%%%%%%%%%%%%%%%%%%%%%%%%%%%%%%%%%%%%%%%%%%

%%%%%%%%%%%%%%%%%%%%%%%%%%%%%%%%%%%%%%%%%%%%%%%%
%%%%%%%%            Sections            %%%%%%%%
%%%%%%%%%%%%%%%%%%%%%%%%%%%%%%%%%%%%%%%%%%%%%%%%
%%%%%%%%%%%%%%%%%%%%%  END %%%%%%%%%%%%%%%%%%%%%
%%%%%%%%%%%%%%%%%%%%%%%%%%%%%%%%%%%%%%%%%%%%%%%%

%%%%%%%%%%%%%%%%%%%%%%%%%%%%%%%%%%%%%%%%%%%%%%%%
 %%%%%%%%          Statements          %%%%%%%%
%%%%%%%%%%%%%%%%%%%%%%%%%%%%%%%%%%%%%%%%%%%%%%%%
\section*{CRediT authorship contribution statement}

\textbf{Jose M. Folgueiras:} Conceptualization, Methodology, Software, Investigation, Formal analysis, Writing – Original Draft, Visualization, Writing – Review \& Editing.  
\textbf{Jorge R. Torga:} Conceptualization, Supervision, Writing – Review \& Editing, Resources.  
\textbf{Eneas N. Morel:} Software, Resources. 
\textbf{Lucas Gabriel Chej:} Software, Validation, Formal analysis, Writing – Original Draft, Visualization.
\textbf{Luis Luciano Zurdo:} Software, Validation, Formal analysis, Visualization. 
\textbf{Alejandro Gabriel Monastra:} Resources, Supervision.  
\textbf{María Florencia Carusela:} Conceptualization, Resources, Supervision, Writing – Review \& Editing.  

\section*{Acknowledgements}  

The authors sincerely thank Pablo Tabla (Electrical Engineer and PhD student, GFA research group) for his expert advice, critical feedback, and hands-on support during the experimental setup development.We also thank CSC-CONICET for providing access to the TUPAC cluster to perform the numerical simulations.

\section*{Funding}

This work was supported by the Universidad Tecnológica Nacional [PID CCTCADE0008423TC and PID MAUTIDE0005320TC]; Universidad Nacional de General Sarmiento [CyTUNGS 30/1161]; and the Consejo Nacional de Investigaciones Científicas y Técnicas (CONICET) [PIP-CONICET 11220200101599CO].

\section*{Declaration of Competing Interest}

The authors declare that they have no known competing financial interests or personal relationships that could have appeared to influence the work reported in this paper.

\section*{Data Availability}

The datasets generated during and/or analyzed during the current study are available from the corresponding author upon reasonable request. Researchers interested in accessing the data for verification or further analysis are encouraged to contact the author directly for detailed discussions and support.

% --- Agregar esto JUSTO ANTES de la seccion de referencias
\section*{Declaration of generative AI in scientific writing}

During the preparation of this work, the authors used ChatGPT and DeepSeek to improve language clarity, grammar, and readability. After using these tools, the authors reviewed and edited the content as needed and takes full responsibility for the content of the publication.

%%%%%%%%%%%%%%%%%%%%%%%%%%%%%%%%%%%%%%%%%%%%%%%%
 %%%%%%%%          Statements          %%%%%%%%
%%%%%%%%%%%%%%%%%%%%%%%%%%%%%%%%%%%%%%%%%%%%%%%%
%%%%%%%%%%%%%%%%%%%%%  END %%%%%%%%%%%%%%%%%%%%%
%%%%%%%%%%%%%%%%%%%%%%%%%%%%%%%%%%%%%%%%%%%%%%%%

%%%%%%%%%%%%%%%%%%%%%%%%%%%%%%%%%%%%%%%%%%%%%%%%
%%%%%%%%       Appendix Section         %%%%%%%%
%%%%%%%%%%%%%%%%%%%%%%%%%%%%%%%%%%%%%%%%%%%%%%%%

%%%%%%%%%%%%%%%%%%%%%%%%%%%%%%%%%%%%%%%%%%%%%%%%
%%%%%%%%       Appendix Section         %%%%%%%%
%%%%%%%%%%%%%%%%%%%%%%%%%%%%%%%%%%%%%%%%%%%%%%%%
%%%%%%%%%%%%%%%%%%%%%  END %%%%%%%%%%%%%%%%%%%%%
%%%%%%%%%%%%%%%%%%%%%%%%%%%%%%%%%%%%%%%%%%%%%%%%

%%%%%%%%%%%%%%%%%%%%%%%%%%%%%%%%%%%%%%%%%%%%%%%%
%%%%%%%%         CITES Section          %%%%%%%%
  \bibliographystyle{elsarticle-num} 
  \bibliography{references_1st_revision}
%%%%%%%%%%%%%%%%%%%%%%%%%%%%%%%%%%%%%%%%%%%%%%%%
%%%%%%%%         CITES Section          %%%%%%%%
%%%%%%%%%%%%%%%%%%%%%%%%%%%%%%%%%%%%%%%%%%%%%%%%
%%%%%%%%%%%%%%%%%%%%%  END %%%%%%%%%%%%%%%%%%%%%
%%%%%%%%%%%%%%%%%%%%%%%%%%%%%%%%%%%%%%%%%%%%%%%%
\end{document}